\documentclass[runningheads]{svmult}

\usepackage{makeidx}   
\usepackage{graphicx}  
\usepackage{subeqnar}  
\usepackage{multicol}  
\usepackage{physprbb}  
\makeindex             

\begin{document}
\title*{Chemical abundances of RGB-Tip stars\protect\newline 
in the Sagittarius dwarf Spheroidal galaxy}
\toctitle{Chemical abundances of RGB-Tip stars\protect\newline 
in the Sagittarius dwarf Spheroidal galaxy}
%
%
\titlerunning{RGB-Tip stars in the Sgr dSph}
%
\author{Lorenzo Monaco}
\authorrunning{Lorenzo Monaco}
%
%
\institute{INAF - Osservatorio Astronomico di Trieste, 
via Tiepolo 11, 34131 Trieste, Italy}

\maketitle              

\begin{abstract}

We present preliminary iron abundances and $\alpha$ element (Ca, Mg) abundance
ratios for a sample of 22 Red Giant Branch (RGB) Stars  in the Sagittarius
galaxy (Sgr), selected near the RGB-Tip. The sample is
representative of the Sgr dominant population. The mean iron abundance is
[Fe/H]=-0.49. The $\alpha$ element abundance ratios are slightly subsolar, in
agreement with the results recently prensented by \cite{boni}. 

\end{abstract}

\section{Target Selection and abundance analysis}

The Sagittarius dwarf Spheroidal galaxy (Sgr dSph) is currently disrupting
under the strain of the Milky Way (MW) tidal field. The study of the Sgr
chemical composition allows us to study at the same time the star formation
history of a dwarf galaxy and the relevance of the hierarchical merging process
for the formation of large galaxies such as the MW.

In May 2003, we obtained spectra for 24 Sgr stars using the high resolution
spectrograph FLAMES-UVES@VLT. The target selection has been performed using the
2Mass infrared photometry (see, e.g., \cite{tip}) where the Sgr Red
Giant Branch stands out very clearly from the contaminating MW field. In this
way we selected  stars belonging to the Sgr dominant population near the
RGB-Tip. Such a selection turned out to be very efficient: 23 out of 24  stars
are Sgr radial velocity members.

Up to date we derived the iron abundance for 22 stars and alpha element
abundance ratios (Mg, Ca) for 20 stars of the sample.  Temperatures have been
derived from the (V-I) color using the calibration of \cite{alonso} and
E(B-V)=0.14 \cite{ls00}.  Gravities were determined superposing theoretical
isochrones \cite{leo} to the observed optical color magnitude diagram while
microturbolent velocities have been derived minimizing the dependence of the
derived abundance from the measured equivalent width.  A model atmosphere has
been computed for each star using the ATLAS~9 code. Equivalent widths have been
measured by using the standard IRAF task SPLOT and, finally, abundances were
determined using the WIDTH code.

\begin{figure}[http]
\begin{center}
\includegraphics[height=0.6\textheight,width=1.\textwidth]{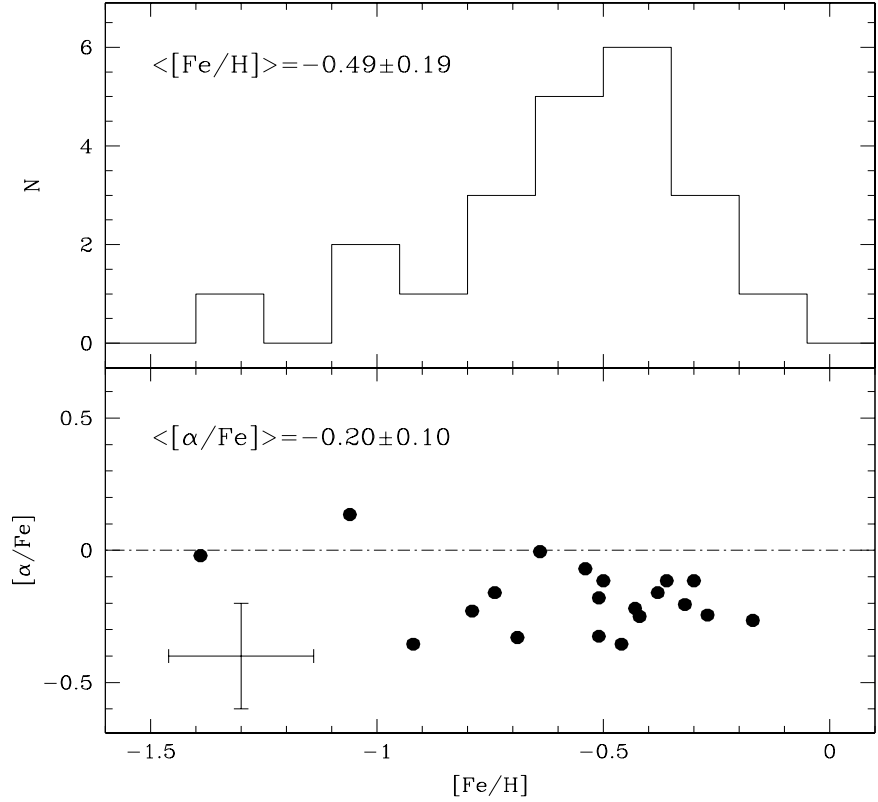}
\end{center}
\caption{Upper panel: metallicity distribution for the 22 stars analyzed so
far. Lower panel: $\alpha$ element abundance ratio {\it vs} iron abundance for 20
stars of the sample. The [$\alpha$/Fe] abundance ratio is the average of [Mg/Fe]
and [Ca/Fe]. A typical errorbar is also plotted.}
\label{eps1}
\end{figure}

As can be seen from the metallicity distribution (see Fig.~\ref{eps1}, upper
panel),  the mean metallicity of the Sgr dominant population is
$<$[Fe/H]$>$=-0.49$\pm$0.19, in excellent agreement with the results derived in
\cite{bump} from the magnitude of the RGB-bump and the shape of the Red Giant
Branch. A component extending  towards lower metallicities is also present. 
The $\alpha$ element abundance ratio (see  Fig.~\ref{eps1}, bottom panel) is
slightly subsolar,  $<$[$\alpha$/Fe]$>$=-0.20$\pm$0.10, in agreement with
\cite{boni} and, apart from a small offset, also with \cite{shmw}.

%

\end{document}